\documentclass[apj]{emulateapj}

\shorttitle{Galaxy Kinematics at $z\sim2$}
\shortauthors{Alcorn et al.}

\usepackage{graphicx} 
\usepackage{color}

\begin{document}

\newcommand{\halpha}{H$\alpha$}
\newcommand{\hbeta}{H$\beta$}
\newcommand{\OII}{[\hbox{{\rm O}\kern 0.1em{\sc ii}}]}
\newcommand{\OIII}{[\hbox{{\rm O}\kern 0.1em{\sc iii}}]}
\newcommand{\NII}{[\hbox{{\rm N}\kern 0.1em{\sc ii}}] }
\newcommand{\zfire}{{\sc ZFIRE}}
\newcommand{\vrot}{$v_{rot}$}
\newcommand{\sint}{$\sigma_{\rm int}$}
\newcommand{\Msun}{M$_{\odot}$}
\newcommand{\Mstar}{M$_{\star}$}
\newcommand{\Mdyn}{M$_{\rm vir}$}
\newcommand{\Mbar}{M$_{\rm baryon}$}
\newcommand{\kms}{km~s$^{-1}$}



\title{ZFIRE: The Kinematics of Star-Forming Galaxies as a Function of
Environment at $\MakeLowercase{z}\sim2$} 


\author{Leo Y. Alcorn\altaffilmark{1}, 
Kim-Vy H. Tran\altaffilmark{1},
Glenn G. Kacprzak\altaffilmark{2},
Themiya Nanayakkara\altaffilmark{2},
Caroline Straatman\altaffilmark{3}, 
Tiantian Yuan\altaffilmark{4}, 
Rebecca J. Allen\altaffilmark{2,5},
Michael Cowley\altaffilmark{5,6},
Romeel Dav{\'e}\altaffilmark{7, 8, 9},
Karl Glazebrook\altaffilmark{2}, 
Lisa J. Kewley\altaffilmark{4}, 
Ivo Labb{\'e}\altaffilmark{3},
Ryan Quadri\altaffilmark{1,10},
Lee R. Spitler\altaffilmark{5,6}, 
Adam Tomczak\altaffilmark{11}}

\altaffiltext{1}{George P. and Cynthia W. Mitchell Institute for Fundamental Physics and Astronomy, Department of Physics \& Astronomy, Texas A\&M University, College Station, TX 77843}
\altaffiltext{2}{Swinburne University of Technology, Hawthorn, VIC 3122, Australia}
\altaffiltext{3}{Leiden Observatory, Leiden University, P.O. Box 9513, NL 2300 RA Leiden, The Netherlands}
\altaffiltext{4}{Research School of Astronomy and Astrophysics, The Australian National University, Cotter Road, Weston Creek, ACT 2611, Australia}
\altaffiltext{5}{Australian Astronomical Observatory, PO Box 915, North Ryde, NSW 1670, Australia}
\altaffiltext{6}{Department of Physics and Astronomy, Faculty of Science and Engineering, Macquarie University, Sydney, NSW 2109, Australia}
\altaffiltext{7}{University of the Western Cape, Bellville, Cape Town, 7535, South Africa}
\altaffiltext{8}{South African Astronomical Observatories, Observatory, Cape Town, 7925, South Africa}
\altaffiltext{9}{African Institute for Mathematical Sciences, Muizenberg, Cape Town, 7945, South Africa}
\altaffiltext{10}{Mitchell Astronomy Fellow}
\altaffiltext{11}{Department of Physics, University of California, Davis, CA, 95616, USA}
\altaffiltext{12}{zfire.swinburne.edu.au}


\begin{abstract}

We perform a kinematic analysis of galaxies at $z\sim2$ in the COSMOS
legacy field using near-infrared (NIR) spectroscopy from Keck/MOSFIRE
as part of the ZFIRE survey. Our sample consists of 75 Ks-band selected star-forming galaxies from the ZFOURGE survey with stellar masses ranging from log(\Mstar/\Msun)$=9.0-11.0$, 28 of which are members of
a known overdensity at $z=2.095$. We measure \halpha\ emission-line
integrated velocity dispersions (\sint) from 50$-$230~\kms, consistent
with other emission-line studies of $z\sim2$ field galaxies. From these data we estimate virial, stellar, and gas masses and derive correlations between these properties for cluster and field galaxies at $z\sim2$. We find evidence that baryons dominate within the central effective radius.  However, we find no statistically significant differences between the cluster and the field, and conclude that the kinematics of star-forming galaxies at $z\sim2$ are not significantly different between the cluster and field environments.

\end{abstract}



\section{Introduction}

At $z>1$, cluster galaxies have significant ongoing star formation \citep{rettura10, tran10, brodwin13, santos14}. The presence of emission lines in cluster galaxies at $z>1.5$ provides an opportunity to investigate the effect of environment on emission line scaling relations. Galaxy properties in the local universe depend strongly on environment, e.g. stellar mass, gas fraction, morphology, and star formation rate (SFR) \citep{dressler80}. However, at $z\sim2$ little evidence for environmental effects on SFR and the Mass-Metallicity Relation \citep{tran10,kacprzak15} and minor effects on size \citep{allen15} have been observed. Kinematics and dynamical masses, which probe more fundamental properties of galaxies, so far have not been tested in cluster environments at $z>1.5$.

Kinematic scaling relations track how mass and luminosity are correlated and can be interpreted in terms of stellar mass and dynamical (total) mass. Studies of local emission line scaling relations, like the Tully-Fisher relation \citep[][TFR]{tullyfisher}, find that cluster and field populations follow the same trends \citep{mocz12, bosch13}. It is unknown if environment is correlated with kinematics at higher redshifts, as few clusters have been confirmed at $z>1.5$. 

Observations of field galaxies show that stellar-mass scaling relations stay relatively consistent with local measurements until $z\sim1.7$ \citep{kassin07, miller11,teodoro16}. Some observations also suggest that these relations evolve at $z>2$ \citep{cresci09, gnerucci11, straatman}. This is possibly because gas fractions are higher at these redshifts, as supported by recent observations \citep{daddi10, tacconi13} and predicted by simulations \citep{dutton11}. It is unclear whether cluster galaxies follow the same trends as the field, or if they evolve at higher redshift. As such, offsets in kinematics between cluster and field galaxies could indicate different evolutionary states in denser environments, e.g. the increasing fraction with redshift of post-starburst galaxies in clusters that span a range of velocity dispersions (Tran et al. 2003).

\begin{figure*}[t]
\centering
\includegraphics[width=0.8\textwidth]{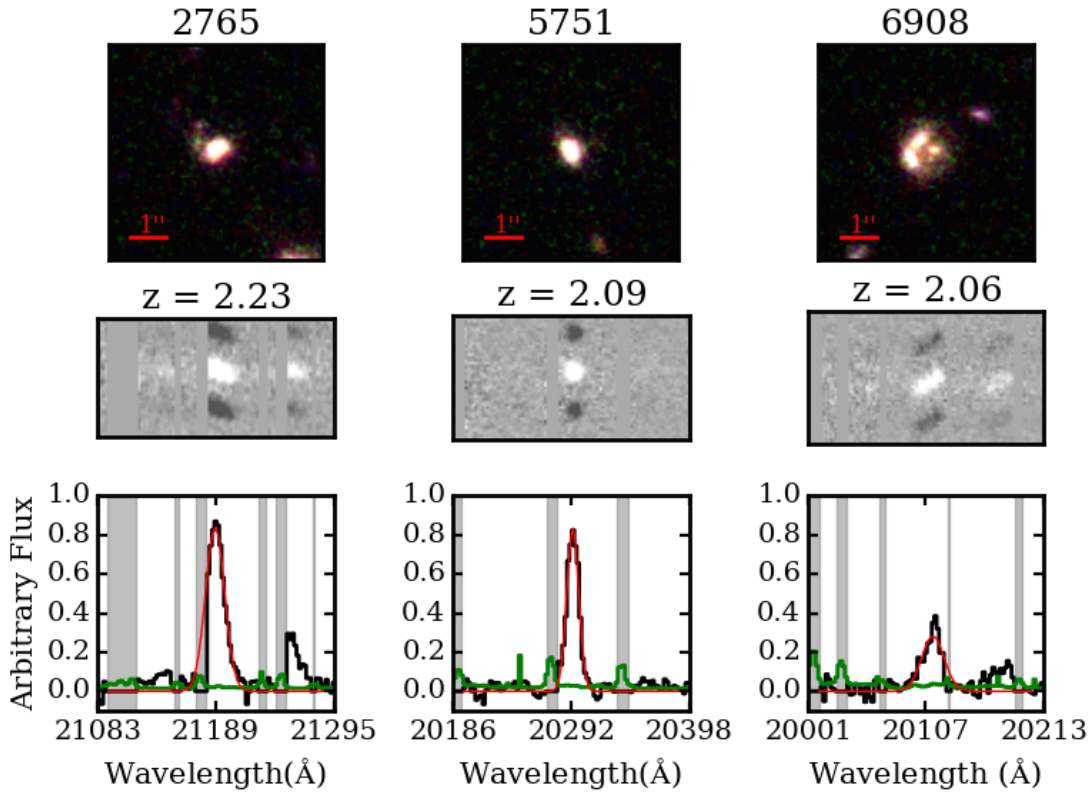}
\caption[width=0.7\textwidth]{Top: RGB images of three galaxies in the ZFIRE sample. ID numbers are object IDs listed in the ZFIRE catalog \citep{themiya}. RGB colors are from CANDELS/3D-HST imaging in F160W (Red), F140W (Green), and F125W (Blue). Middle: Example flux and telluric corrected spectra from the ZFIRE pipeline corresponding to the images in the top row. Here we see the \halpha\ and \NII  emission lines. Bottom: 1D summed spectra in black, and the error spectrum in green. We plot the Gaussian fit to the data in red. Masked sky regions are in grey.}
\end{figure*}

Observations have shown that kinematics for both resolved and unresolved objects can be tracked using integrated velocity dispersion, $\sigma_{int}$, measured with emission lines such as \halpha\ (for a review of kinematic surveys using this technique, see \citet{glazebrook13}). Here we present the most distant study yet to compare \halpha\ kinematics of individual cluster galaxies and field galaxies. Our data consists of objects measured by the ZFIRE\altaffilmark{12} survey \citep{themiya}, including the $z=2.095$ overdense region in the COSMOS field \citep{spitler12,yuan14}.

 ZFIRE targets galaxy clusters at $z\sim2$ to explore galaxy evolution as a function of environment.  ZFIRE combines deep multi-wavelength imaging with spectroscopy obtained from Keck/MOSFIRE \citep{mclean12} to measure galaxy properties including sizes, stellar masses, star formation rates, gas-phase metallicities, and the interstellar medium \citep{kacprzak15,kewley15,tran15,themiya}.

In this work, we assume a flat $\Lambda$CDM cosmology with $\Omega_{M}$=0.3, $\Omega_{\Lambda}$ =0.7, and H$_{0}$=70. At the cluster redshift, $z = 2.09$, one arcsecond corresponds to an angular scale of 8.33 kpc.

\section{Data}

\subsection{HST Imaging}

Our morphological measurements are from Cosmic Assembly Near-Infrared Deep Extragalactic Survey imaging \citep[][CANDELS]{koekemoer, grogin11} processed by the 3D-HST team (v4.1 data release). For details on the reduction of CANDELS imaging, see \citet{skelton14}. We use GALFIT software \citep{peng10} to measure galaxy sizes from the F160W imaging. Examples of CANDELS/3D-HST imaging of galaxies in our sample can be seen in Figure 1.

We generate a custom pipeline to fit the 161 COSMOS galaxies in ZFIRE with F160W imaging using initial measurements of size, axis ratio ($q$), position angle (PA), and magnitude from SExtractor \citep{banda96}. Our constraints are adopted from the constraints in \citet{vdw12}, and our point-spread function is constructed by 3D-HST. Objects within 2$\arcsec$ of a target galaxy are simultaneously fit with the central object. Residual images are visually inspected to determine the best possible fits for each galaxy. Galaxies with poor residuals are refit using a different set of initial parameters, and rejected if a satisfactory solution can not be obtained. Our results are consistent within $2\sigma$ to \citet{vdw12}.

\begin{figure}[h]
\plotone{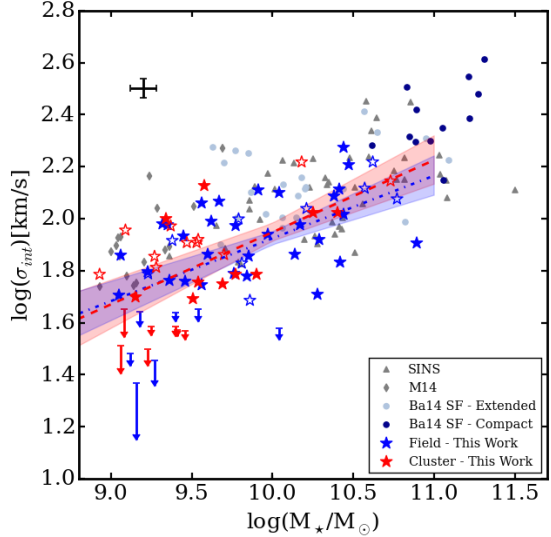}
\caption[width=0.7\textwidth]{Stellar mass vs $\sigma_{int}$ of the ZFIRE galaxies. Size and color of the points refers to the quality of the spectroscopic measurement: filled points for confirmed line widths, outlined points for faint emission lines or emission lines partially obscured by sky interference. Blue stars are field galaxies, and red stars are galaxies in the $z\sim2.09$ cluster identified in \citet{spitler12}. Characteristic errorbars are located in the upper left, in black. We compare the ZFIRE COSMOS sample with emission-line $z\sim2$ field galaxies from \citet{barro14} (written as Ba14), \citet{sins} (written as (SINS), and \citet{masters14} (written as M14). Objects with line width less than instrumental resolution are displayed as upper limits. The bootstrapped 1$\sigma$ confidence intervals of the least-squares linear fits are shown as translucent boxes around the best fit lines. We see no significant difference between the best-fit relations for cluster and field.}
\end{figure}

Errors for our GALFIT measurements are obtained by adding sky noise to the GALFIT model, and rerunning GALFIT 200 times per object. The range of the error is obtained from the 1$\sigma$ confidence intervals.

The size of the galaxy is obtained using $q$ and the effective radius, $r_{e}$, from GALFIT. We convert this to a circularized effective radius using $R_{e}=r_{e}\sqrt{q}$. The properties of our overall population show no significant size dependence on environment. This is in conflict with \citet{allen15} who find evidence that star-forming galaxies in the cluster are larger than in the field.  However, we stress that this is likely because our analysis is limited to the smaller set of \halpha$-$detected galaxies.

\subsection{ZFOURGE Photometry}

The COSMOS cluster was initially identified by \citet{spitler12} using photometric redshifts from ZFOURGE \citep{straatmanzfourge} and subsequently confirmed with spectroscopic redshifts from MOSFIRE \citep{yuan14}. ZFOURGE combines broad-band imaging in K and the medium-band J$_{1}$, J$_{2}$, J$_{3}$, H$_{s}$, and H$_{l}$ filters to select objects using Ks-band images with a 5$\sigma$ limit of 25.3 AB magnitudes.

ZFOURGE uses FAST \citep{kriek09} to fit stellar population synthesis models to the galaxy spectral energy distributions to estimate observed galaxy properties. We assume a \citet{imf} initial mass function with constant solar metallicity and an exponentially declining star formation rate, and a \citet{calzetti} dust law. 

\subsection{MOSFIRE NIR Spectroscopy}

Observations were taken in December 2013 and February 2014 in the K-band filter covering 1.93-2.45 $\mu$m, the wavelength range we would expect to see \halpha\ and \NII at the cluster redshift. Targets were star-forming galaxies (SFGs) selected from rest frame UVJ colors. Seeing varied from  $\sim0.4\arcsec$ to  $\sim0.7\arcsec$ over the course of our observations.

\begin{figure*}[t]
\centering
\includegraphics[width=0.9\textwidth]{f3.pdf}
\caption[width=0.7\textwidth]{Left: Estimated virial masses obtained from the virial formula vs. stellar masses. Best-fit relations are included with 1$\sigma$ deviation boxes, in the same colors as represented in Figure 2. Right: Virial masses binned by stellar mass. Errors in  log(\Mstar/\Msun) are the width of each bin. We compare the ZFIRE sample to the $z\sim2$ field galaxies of \citet{sins}, \citet{barro14}, and \citet{masters14} to find consistent values with extended SFGs.}
\end{figure*}

The spectra are flat-fielded, wavelength calibrated, and sky subtracted using the MOSFIRE data reduction pipeline (DRP)\footnote[1]{http://www2.keck.hawaii.edu/inst/mosfire/drp.html}. We use a custom ZFIRE pipeline to correct for telluric absorption and perform a spectrophotometric flux calibration using a type A0V standard star.  We flux calibrate our objects to the continuum of the standard star, and use ZFOURGE photometry to correct offsets between photometric and spectroscopic magnitudes. The final result of the DRP are flux-calibrated 2D spectra (see examples in Figure 1) and 2D 1$\sigma$ images used for error analysis, with a bootstrapped flux calibration error of $<10\%$ ($\sim$0.08 magnitudes). For more information on ZFIRE spectroscopic data reduction, see \citet{themiya}.

We extract 1D spectra from an aperture the width of the one Gaussian sigma ($1\sigma$) boundaries of the spatial \halpha\ emission-line profile. Varying the aperture width does not affect our results. The 1D \halpha\ line width is determined by fitting a Gaussian profile to the \halpha\ emission line. We subtract the measured instrumental broadening in quadrature from the line width, and convert the corrected line width to \sint\ using the best-fit redshift from \citet{themiya}. Errors are calculated by adding sky noise to the observed spectrum, and refitting 1000 times.

We test if slit misalignment affects our results. After rejecting objects with $\Delta\alpha$ $>$ $40^{\circ}$, where $\Delta\alpha$ is the difference between the GALFIT-measured PA and the angle of the slit, our sample decreases to 26 objects overall, 7 in the cluster. However, our results do not significantly change, so we conclude the slit misalignment does not significantly affect our final results for our scaling relations or virial mass measurements and do not include galaxy PA corrections or restrictions in our analysis.

All Gaussian line fits are visually inspected. Emission lines with sky contamination are given a lower quality flag than emission lines without contamination, but are included in our sample (Figure 2). Measurements with signatures of AGN as detected in \citet{cowleysub}, objects completely obscured by sky emission, or objects too faint to detect manually are excluded from our analysis. After our rejection criteria, the sample contains 75 COSMOS galaxies, 28 of which are associated with the $z\sim2$ cluster.

The cluster objects are defined as objects identified with three strongly overdense regions of the COSMOS field. In \citet{spitler12}, these overdensities are found by computing surface density maps. In \citet{yuan14}, these objects are spectroscopically confirmed and concentrated at $z_c=2.095$, and are consistent with a Gaussian distribution with $\sigma_z=0.005$. The redshift range for the cluster is defined to be $z_c\pm3\sigma_z$. The COSMOS overdensity velocity dispersion is measured to be $\sigma_{v1D}=552\pm52$ km s$^{-1}$, and has 57 spectroscopically confirmed members. It consists of four major groups that cover a total projected size
of $3.7\times5$ Mpc$^2$ ($7.4 \times 10$ Mpc$^2$ comoving). The cluster is most likely to evolve into a Virgo-like cluster at $z\sim0$ \citep[from $\Lambda$CDM simulations, see][]{yuan14}. Field objects are defined as targetted objects not within the cluster redshift range or associated spatially with the cluster.

\section{Results}

\subsection{\halpha\ Emission-Line Widths at $z\sim2$}

By measuring kinematics from the \halpha\ line width, we assume that the broadening is caused by the gravitational potential of the galaxy acting on the gas. We also use \sint\ since it can be measured for all galaxies, even those with unresolved rotation, and is robust against PSF effects. \sint\ could trace rotation, velocity dispersion, or a combination of both quantities \citep{glazebrook13,barro14,masters14}. \halpha\ integrated velocity dispersions of ZFIRE COSMOS galaxies range from $\sim$50-230 km s$^{-1}$, an expected distribution of values for extended (rather than compact) SFGs (Figure 2). The median \sint\ is 72.8 km s$^{-1}$.

We determine a linear least-squares fit to the cluster, field, and total ZFIRE samples (Table 1) normalized at log(\Mstar)=10. The least-square linear fits relation is of the form log$(y)$ = $A$(log$(x)$-10) + $B$. The best-fit relation is bootstrapped 1000 times to determine 1$\sigma$ confidence intervals of the linear fit. The best-fit log(\Mstar)-log(\sint) relations for cluster ($A=0.28\pm0.06$, $B=1.95\pm0.03$) and field ($A=0.24\pm0.05$, $B=1.92\pm0.03$) are consistent within 1$\sigma$ (Table 1), indicating no evidence of environmental influence on kinematics.
The median residual of the points around each best-fit line is $\sim0.12$ dex, and cluster and field relations overlap within this scatter.  Our results do not depend on whether or not we apply weighting from our errors on \sint.

To quantify our ability to recover 1D line widths, we use a set of 2D emission-line models with exponential disks, an arctan rotation profile, and known $v_{rot}$ and gas $\sigma$. We add sky noise (as measured from our data) to the modeled galaxies, and collapse each emission line (simulated and with sky noise) to 1D line widths. We find that the simulated emission lines with noise differ by only $\sim$0.01\% compared to the input models. 

We confirm that our results on cluster versus field do not depend on inclination corrections. To correct for inclination, \citet{straatman} and \cite{price15} assume that the intrinsic axis ratio is $q_0=0.19$.  When we apply this correction, the scatter of our points around our best-fit values decreases by $\sim0.01$ dex and the values are offset from the uncorrected values by $0.05$ dex.  If we do not correct for inclination, we tend to underestimate the input virial mass of our modeled galaxies by $\sim0.25$ dex.  However, this assumes that the models accurately represent the true galaxy kinematics.  Because an inclination correction does require assuming an intrinsic axis ratio and applying an inclination correction does not change our overall results, we use uncorrected \sint\ values so that we can  compare directly to recent results by \citet{barro14,masters14}.  We will explore the effects of inclination corrections in future work.

\begin{figure*}[t]
\centering
\includegraphics[width=0.9\textwidth]{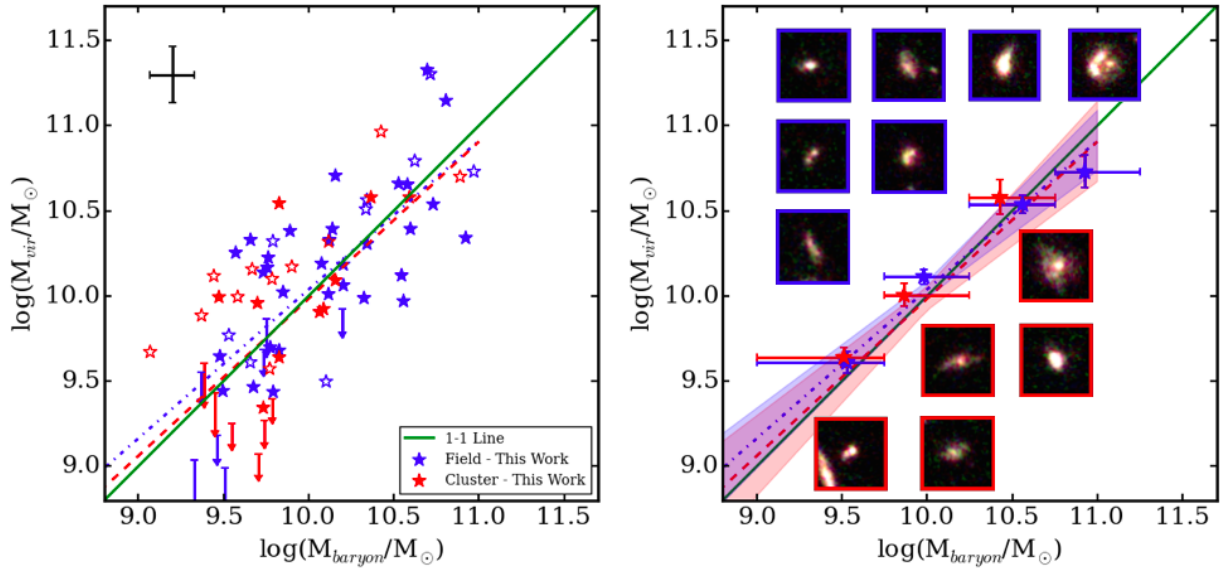}
\caption[width=0.7\textwidth]{Left: Virial masses obtained from the virial formula compared to estimated baryonic masses. Baryonic masses are from the addition of stellar masses and gas masses computed from dust-corrected \halpha\ fluxes and the Kennitcutt-Schmidt relation \citep{kennicutt}. We use the same colors as represented in Figure 2, with characteristic errorbars for our data in black in the upper left of the panel. Right: Virial mass binned by baryonic mass. We include $3\arcsec \times 3\arcsec$ RGB images of selected galaxies in each bin. Borders around the RGB cutouts are blue for field galaxies, and red for cluster galaxies.}
\end{figure*}

In the right panel of Figure 1 we compare the ZFIRE sample to the 1D field objects of \citet{barro14}, the 1D field objects of \citet{masters14}, and the 2D Integral Field Unit (IFU) field objects of the Spectroscopic Imaging survey in the NIR with SINFONI sample \citep[][SINS]{sins}. We see consistent values of \sint\ with SFGs in these samples.
We also have compared our emission-line kinematics with the independent analysis of ZFIRE spectra presented by \citet{straatman}, and find consistent 1D line widths with their collapsed best-fit 2D kinematic models.

\subsection{Virial Masses}

In Figure 3 we compare estimated virial mass to stellar mass for the COSMOS field and cluster objects and to other high-redshift kinematic surveys. While stellar mass can be estimated by spectral energy distribution fitting, which examines stellar populations and colors, virial mass accounts for the total mass of the galaxy including gas and dark matter.

To calculate virial mass, we apply the virial formula 
\begin{equation}
M_{\rm vir}(<R_e)=K_{e}\frac{\sigma_{int}^2R_{e}}{G}.
\end{equation}
For comparison to recent results by \citet{barro14, masters14}, we assume a virial factor K$_{e}=5$.  We find that our virial masses are consistent with these existing studies and \citet{sins}.  Our linear best-fit relations for the cluster and field (Table 1) shows no significant environmental impact on our fits: cluster and field relations differ by $<1\sigma$.

In our anaysis, we use a constant virial factor but note that $K_{e}$ depends on the structural parameters of the galaxy.  Values in the literature range from $K_{e}\sim 2-10$ and using, $e.g.$ K$_{e}=3.4$ from \citet{erb06} will introduce an offset of $\sim-0.2$ dex in our results.  However, we stress that the relative comparison between cluster and field does not change.

\subsection{Estimating Gas Mass from the Kennicutt-Schmidt Relation}

Gas masses are estimated from the Schmidt-Kennicutt relation (referred to here as the Kennicutt-Schmidt relation) \citep{kennicutt}, using SFRs from \halpha\ line fluxes corrected for dust using the method described in \citet{tran15}. We assume a nebular attenuation from \citet{cardelli} with R$_{V}=3.1$, and a \citet{calzetti} dust law. We calculate the SFR surface density, $\Sigma_{SFR}=SFR/(\pi R_{H\alpha}^{2})$ and use the Kennicutt-Schmidt relation to solve for gas surface density, $\Sigma_{gas}=M_{gas}/(\pi R_{H\alpha}^{2})$. $R_{H\alpha}$ is the circularized effective radius in kiloparsecs, after we have applied a correction factor from \citet{nelson15} to estimate the \halpha\ radius from the stellar radius and mass. We add the estimated gas mass to the stellar mass to derive baryonic masses.

\begin{deluxetable*}{ccccccc}
\tabletypesize{\scriptsize}
\tablecaption{Values for all weighted least-square linear fits\tablenotemark{a} to ZFIRE cluster and field data}
\tablewidth{0pt}
\tablehead{
\colhead{$x$} & \colhead{$y$} & \colhead{Environment} & \colhead{$A$\tablenotemark{b}} & \colhead{$B$\tablenotemark{b}} &\colhead{$N$\tablenotemark{c}}&\colhead{Residual\tablenotemark{d}} }
\startdata
\Mstar & \sint\tablenotemark{e} & Cluster & 0.28$\pm$0.06 & 1.95$\pm$0.03 & 28 & 0.12 \\
{ } & { } & Field & 0.24$\pm$0.05 & 1.92$\pm$0.03 & 47 & 0.13 \\
{ } & { } & Total & 0.25$\pm$0.04 & 1.93$\pm$0.02 & 75 & 0.11 \\
\Mstar & \Mdyn & Cluster & 0.86$\pm$0.16 & 10.23$\pm$0.08 & 28 & 0.30 \\
{ } & { } & Field & 0.79$\pm$0.13 & 10.26$\pm$0.06 & 47 & 0.21 \\
{ } & { } & Total & 0.82$\pm$0.10 & 10.25$\pm$0.05 & 75 & 0.24 \\
\Mbar & \Mdyn & Cluster & 0.92$\pm$0.17 & 9.98$\pm$0.07 & 28 & 0.32 \\
{ } & { } & Field & 0.87$\pm$0.12 & 10.04$\pm$0.05 & 47 & 0.23 \\
{ } & { } & Total & 0.90$\pm$0.11 & 10.02$\pm$0.04 & 75 & 0.25 
\enddata
\tablenotetext{a}{The least-square linear fits relation is of the form log$(y)$ = $A$(log$(x)$-10) + $B$.}
\tablenotetext{b}{Errors are determined by bootstrapping the data 1000 times, and determining the 1$\sigma$ confidence intervals of the bootstrapped results.}
\tablenotetext{c}{$N$ is the number of objects used for the linear fit.}
\tablenotetext{d}{The residual quoted is the median residual value from the best fit line.}
\tablenotetext{e}{\sint\ is in units of km s$^{-1}$. We do not apply a weight to these fits.}
\end{deluxetable*}

In Figure 4, we show a comparison between estimated baryonic and virial masses. Objects move closer to the unity relation than in Figure 3 due to high gas masses (the median gas fraction in the ZFIRE sample is 0.36, typical of the field galaxies seen in \citet{tacconi13}). This implies most objects are baryon-dominated within one effective radius. We again find no significant environmental impact on the values for \Mbar\ vs. \Mdyn, relations are consistent within 1$\sigma$. The MOSDEF survey similarly finds that field galaxies at $z\sim2$ are distributed around the 1-1 line \citep{price15}.

\section{Summary}

Using the Keck I MOSFIRE NIR spectrograph, we measure \halpha\ emission lines of 28 COSMOS $z=2.095$ star-forming cluster galaxies and 47 star-forming field galaxies to investigate environmental effects on high-redshift protoclusters. Our objects are rest frame UVJ selected star-forming galaxies with no detected X-Ray, IR, or radio AGN signatures. We measure \halpha\ line widths to derive integrated velocity dispersions, $\sigma_{int}$, and use CANDELS/3DHST F160W imaging to measure galaxy sizes. 

We derive high-redshift emission-line kinematic scaling relations and did not find any significant environmental effects; cluster and field least-squares linear relations were consistent within error. Compared to previous multi-slit and IFU kinematic surveys of the $z\sim2$ field, \citet{barro14, masters14, sins}, we find consistent results in the \Mstar-\sint\ relation with extended SFGs.

We estimate virial masses (which includes stellar, dark matter, and gas mass) for our galaxies from our gas kinematics. Gas masses were derived from dust-corrected \halpha\ star-formation rates and the Kennicutt-Schmidt relation, and added to stellar masses to estimate baryonic masses. The median values for log(\Mdyn/\Mstar) and log(\Mdyn/\Mbar) are 0.30 and 0.07, respectively. We find consistent values between baryonic and virial mass estimates, showing baryon dominance within one effective radius from the center of these galaxies. There is no statistically significant evidence of environmental impacts on our sample; cluster and field best-fit relations in \Mstar-\Mdyn\ and \Mbar-\Mdyn\ are consistent within 1$\sigma$.

Our results demonstrate that the integrated gas kinematics of SFGs in the $z=2.095$ overdensity are not strongly dependent on environment. Further studies of $z>1.5$ cluster galaxies are needed to confirm our results. In addition, studies of $z>1.5$ cluster galaxy absorption-line kinematics would also provide an opportunity to compare gas and stellar kinematics. In future work we will present an analysis of the Tully-Fisher relation of our galaxies to investigate the contributions of rotational velocity and velocity dispersion as a function of environment.

\acknowledgments

We are grateful to the anonymous referee for a helpful report. We thank G. Barro for the use of his data, and S. Price, M. Kriek, Jimmy, and J. Walsh for helpful discussions. This work was supported by a NASA Keck PI Data Award administered by the NASA Exoplanet Science Institute. Data presented herein were obtained at the W. M. Keck Observatory from telescope time allocated to NASA through the agency's scientific partnership with the California Institute of Technology and the University of California. This work is supported by the National Science Foundation under Grant \#1410728.  GGK acknowledges the support of the Australian Research Council through the award of a Future Fellowship (FT140100933). The authors wish to recognize and acknowledge the very significant cultural role and reverence that the summit of Mauna Kea has always had within the indigenous Hawaiian community. We are most fortunate to have the opportunity to conduct observations from this mountain. We wish to thank the Mitchell family, particularly the late George P. Mitchell, for their continuing support of astronomy.

\end{document}